\documentclass[twocolumn]{revtex4-1}  

\usepackage{graphicx}
\usepackage{natbib}
\usepackage{amsmath,amssymb}

\begin{document}
\title{A simple decomposition of European temperature variability capturing the variance from days to a decade}

\author{Philipp G Meyer}
\affiliation{Max Planck Institute for the Physics of Complex Systems,
Noethnitzer Str. 38
D 01187 Dresden,
Germany}
\author{Holger Kantz}
\affiliation{Max Planck Institute for the Physics of Complex Systems,
Noethnitzer Str. 38
D 01187 Dresden,
Germany}

\date{\today}

\begin{abstract}
We analyze European temperature variability from station data with the method of detrended fluctuation analysis. This method is known to give a scaling exponent indicating long range correlations in time for temperature anomalies. However, by a more careful look at the fluctuation function we are able to explain the emergent scaling behaviour by short time relaxation, the yearly cycle and one additional process. It turns out that for many stations this interannual variability is an oscillatory mode with a period length of approximately 7-8 years, which is consistent with results of other methods. We discuss the spatial patterns in all parameters and validate the finding of the 7-8 year period by comparing stations with and without this mode.
\end{abstract}

\maketitle

\section{Introduction}

Variability of temperature time series has been described in two ways. Some people concentrate on scaling behavior of correlations and its seeming power law behaviour (Long range correlations). Popular methods are detrended fluctuation analysis (DFA) \citep{Fra03}, and multifractal versions \citep{Moo18}, power spectral densities \citep{Fre16} and wavelet analysis \citep{Lov18}. Such results consider the complete power of climate fluctuations. They are useful for estimations of prediction errors \citep{Mas16,Lud16} and the validation of more complicated general circulation models \citep{Gov02}.

Other projects have tried to identify characteristic timescales in the data \citep{Ghi02}. Here people use for example singular spectrum analysis (SSA) \citep{Pla95}, versions of Monte Carlo SSA \citep{Pal04}, conditional mean approaches \citep{Jaj16}, and again wavelet methods \citep{Bal97}. Results of these methods also help to validate general circulation models and contribute to the physical understanding of the atmosphere as well as to make predictions if the component is significant like the El Nino southern oscillation \citep{Che11}. They can explain observed slowdown (or fastening) of global warming \citep{Ste15,Hu17}.

The disadvantage of the first approach is that it neglects properties that are already known and reduces the information to one property which is difficult to interpret \citep{Mar04}. The disadvantage of the second approach is that it does not take the full power of the fluctuations into account \citep{Lov15}.

While parts of the literature concentrating on characteristic timescales completely ignore the findings of power law scaling, others mention the so called continuum variability \citep{Huy06} and treat it as one of the features contained in temperature fluctuations. Recently interest in this continuum variability is growing again \citep{Lov18,Fre17}. In \citep{Lov15} the author suggest to filter the known frequency modes, just like it is normally done with the seasonal cycle, in order to understand the nature of the continuum better. Our approach in fact follows this spirit. We will apply a novel method based on DFA that in contrast to traditional usage of DFA concentrates on the characteristic timescales. Therefore we show connections between the two approaches mentioned at the beginning. However, we conclude that all the power of the fluctuations can be explained by a superposition of few simple processes for the time range under consideration.

We want to concentrate on climate data from Europe where a large number of stations with several decades of daily recordings exists. For this region there is already a long history of investigations on both individual stations or grid data. Most importantly, people have repeatedly reported a 7-8 year cycle in various stations and other records \citep{Pla95,Pal14,Sen16}, the origin of which is still unclear. Spacial patterns were described in \citep{Gri02,Pis09,Jaj16}, however, to our knowledge there has not been an algorithmic investigation on a similar number of station as we perform it.

We present the foundations of our method in section 3, summarizing previous work in \citep{Mey19NJP}. In section 4 we postulate our model for weather and macroweather variability in Europe and extend our method to processes with more than one characteristic timescale. In section 5 we show results of a large scale algorithmic application of our method to station data of European temperatures. We are able to reproduce the 7-8 year cycle in part of this data. We show that our method is able to separate data with and without this frequency mode.

\section{Data}
We analyze station data of temperature and pressure in Europe \citep{Tan02}. It can be downloaded from www.ecad.com. The resolution of the time series is one day. The datasets strongly vary in length. We restrict ourselves to long time series as described in appendix \ref{ap3}.

\section{The method}
\begin{figure*}
\includegraphics[width=\linewidth]{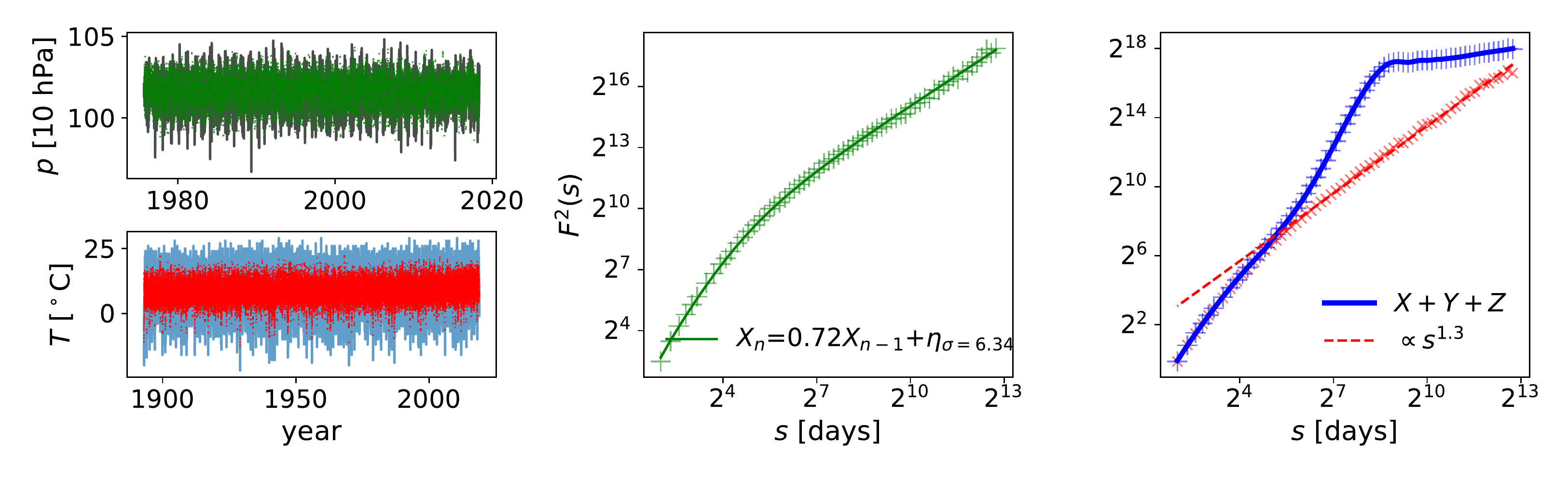}
\caption{LEFT: Pressure (top) and temperature data (bottom) from Potsdam station. The measured values are displayed with solid lines in the background. The anomalies are the dots in the foreground. CENTER: fluctuation function of the pressure anomalies (dots) fitted by an AR(1) model (solid line). The fit indicates a relaxation time of 3 days. RIGHT: fluctuation function of the temperature anomalies (red) fitted by a scaling law of $\alpha=0.65$, and the fluctuation function of the real temperature data (blue) fitted by our model (equation (\ref{TXYZ})).}
\label{fig1}
\end{figure*}

Our method is based on detrended fluctuation analysis (DFA) \citep{Pen94} (see appendix \ref{ap1} for implementation) and its theoretical understanding \citep{Hoe15a}. DFA produces the so called fluctuation function ${F}$ of the time series, which is in fact a nonlinear transformation of the autocorrelation function ${C}$
\begin{equation}
\label{F(C)}
{F}_x^2(s)=\sigma_x^2\left(L_q(0,s)+2\sum_{t=1}^{s-1}C_{xx}(t)L_q(t,s)\right).
\end{equation}
Here $\sigma_x^2$ is the variance of the time series and $L_q$ is a kernel described in appendix \ref{ap2}.
It turns out, that looking at ${F}(s)$ has two advantages over looking at ${C}(t)$: on the one hand it is numerically more stable and on the other hand DFA is able to remove polynomial trends from the signal, so they do not contribute to the fluctuation function \citep{Hoe19}. We only use DFA1, because the trend due to climate change is sufficiently small compared to the fluctuations \citep{Mas16}.

In \citep{Mey19NJP} the authors have shown that this method has the ability to uncover characteristic timescales in time series. The idea is to fit the theoretical fluctuation function of autoregressive models of different orders: AR(1) models for relaxation times, and AR(2) for noisy oscillations. The method works well for data that is dominated by one characteristic timescale, like approximations to yearly global mean temperatures \citep{Mey18JGR} or an 11-month averaged El Nino signal. We show the theoretical fluctuation functions in (\ref{Far1}) and (\ref{Far2}) in appendix \ref{ap2}.

As a first example we want to show the application of the method to the time series of air pressure measured at Potsdam, Germany. We deseasonalize both the pressure and its variance in order to get a homogeneous time series. Therefore we calculate the long time climatological average pressure for each calender day and subtract it from each day. We also divide each value by the average variance for the calender day. 

In the long time limit $F(s)$ scales like $1/2$, implying short range correlations.
Figure \ref{fig1} shows that the data can be described by an AR(1) model in a fairly good approximation. Obviously, for a complex system like the atmosphere, we would not claim that pressure is purely autoregressive and no other effects happen at characteristic timescales. However, we can see that those other timescales are not significant for a smoothed measure such as the DFA-fluctuation function. The method shows us a data model that can be used for stochastic predictions and qualitative understanding of the dynamics.

In figure \ref{fig1} we also show temperatures and temperature anomalies ($T$ with climatological average temperatures for each calender day subtracted) from Potsdam. In DFA the temperature anomalies exhibit anomalous scaling proportional to $s^{0.65}$, which could be described by a model with long range correlations \citep{Mas16}. In this article we want to show a different approach. Instead of looking at the anomalies, we look at $F_T$ itself. We want to identify characteristic timescales and therefore extend our understanding beyond the description of emergent scaling.
If the time series is indeed a superposition of several independent processes we can make use of the superposition principle of DFA \citep{Hu01}
\begin{equation}
\label{FpF}
{F}_{x+y}^2(s)={F}_x^2(s)+{F}_y^2(s).
\end{equation}
This will be important for our purpose in the next section.

\section{Decomposition of one time series}
\begin{figure*}
\includegraphics[width=\linewidth]{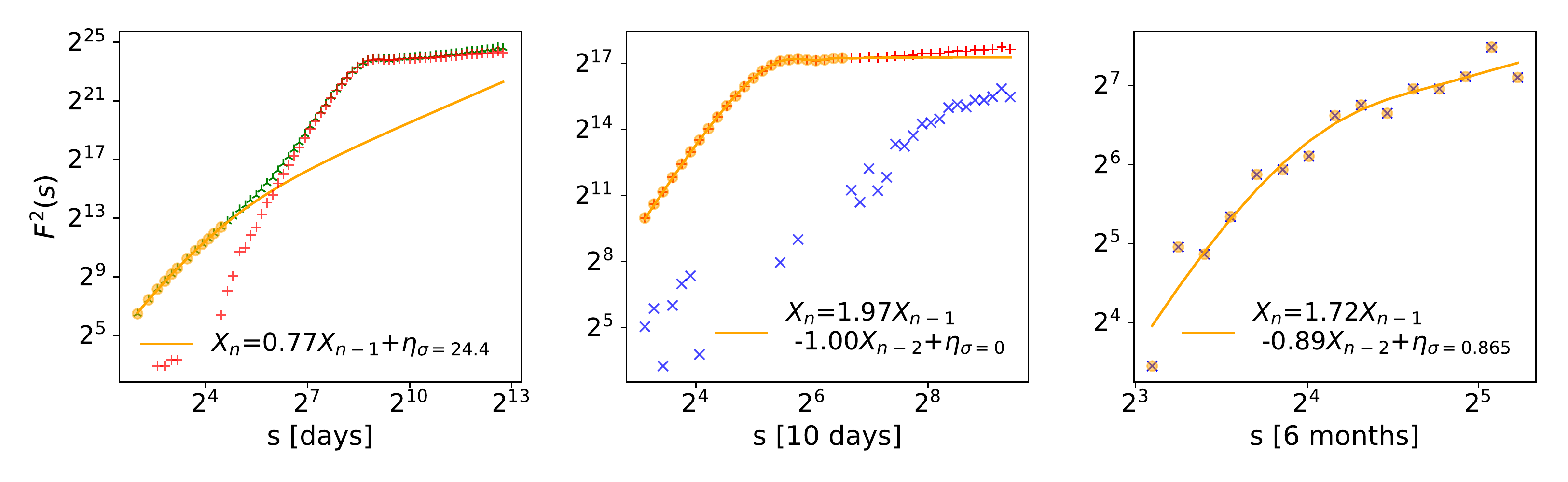}
\caption{Decomposition of the fluctuation function of Potsdam temperatures. The orange curves represent fits to the fluctuation functions. The orange circles mark the points for which the fit is performed. LEFT: short time behaviour $F^2_X$ fitted to the full fluctuation function $F_T$ (green '$\curlywedge$'), which yields a relaxation time of 3.8 days. The red '+' denotes $F^2_{T-X}=F^2_T-F^2_X$; CENTER: seasonal cycle $F^2_Y$, fitted to $F^2_{T-X}$. The resulting $F^2_{T-X-Y}$ is plotted as blue 'x'; RIGHT: interseasonal variability $F^2_Z$ fitted to $F^2_{T-X-Y}$.}
\label{fig2}
\end{figure*}

In the examples above fitting fluctuation functions was exclusively applied to systems with one dominant characteristic timescale. What are we supposed to do if the fluctuation function implies a more complex dynamics? In this section, by analysing historic temperature data from Potsdam station, we show that our method still works for systems with more than one characteristic timescale if these timescales are sufficiently well separated.

We assume that the time series can be written as a superposition of three processes
\begin{equation}
\label{TXYZ}
T_t\approx X_t+Y_t+Z_t,
\end{equation}
namely, a short range part $X_t$, the seasonal cycle $Y_t$ and one unspecified component, which describes interseasonal variability. We will see that these components are clearly separated in time, such that $Y_t$ and $Z_t$ might as well be interpreted as additional components of the noise input $\epsilon_t$ of $X_t$. We will describe $X_t$ and $Z_t$ by simple autoregressive models, while the seasonal cycle $Y_t$ is a regular oscillation.

Temperature correlations for short times decay exponentially \citep{Mas16}. The behaviour is apparent in figure \ref{fig2} in the fluctuation function $F_T(s)$ for small $s$.
Other subseasonal timescales are not pronounced in the fluctuation function. This lack of characteristic timescales is known and the reason why predictions on these timescale are so difficult \citep{Ghi19}. Hence $X_t$ is best described by an autoregressive model of order one AR(1)
\begin{equation}
\label{AR1}
X_{t_X}=cX_{t_X-1}+\epsilon_{t_X} \mbox{ \ \ \ and \ \ \ } r=-1/\log(c),
\end{equation}
where the noise term $\epsilon_{t_X}$ is the output of some process on a much smaller timescale. The relaxation time $r$ is determined by the AR-parameter $c$. The shape of the distribution of $\epsilon_{t_X}$ is not important for our model \citep{Mey19NJP}, however, it seems to be well approximated by a Gaussian distribution. The variance $\sigma^2_\epsilon$ is a free parameter. For simplicity we will later look at the standard deviation $\sigma_X$ of the process instead of $\sigma_\epsilon$. They are related by equation (\ref{AR1V}).

Due to the temporal separation of the three processes $X_t$, $Y_t$ and $Z_t$, and because autoregressive models are low pass filters, the impact of the slower processes can be neglected for short times. For Potsdam the fit of equation (\ref{Far1}) to $F_T$ for small $s$ (see figure \ref{fig2}) yields a relaxation time of 3.8 days. We subtract the fitted AR(1) fluctuation function $F^2_X$ from $F^2_T$.

Since we now covered the fluctuations for short times we change the timescale to 10 days by dividing $F_{T-X}$ and $s$ by 10 and remove small values of $s$. The by far most pronounced mode in the fluctuation function is the seasonal cycle \citep{Den18}. Although it is not perfectly periodic it has a phase-locked frequency of 1/365.25 days, which is enforced by the periodic driving of the sun. We do not remove the cycle by subtracting the climatological average since this popular procedure does not account for the whole dynamics on the yearly timescale. It does not take into account that the phase of the oscillation fluctuates. Instead we approximate the fluctuation function $F_Y$ by an AR(2) process (see equation (\ref{AR2})) with $b=-1$, which is equivalent to a sinusoidal signal
\begin{equation}
\label{SIN}
Y_{t_Y}=\frac{\sigma_Y}{2} \sin\left(\frac{2\pi t_Y}{1 \mathrm{year}}\right),
\end{equation}
which does not describe the phase fluctuations, but accounts for their full power.
Figure \ref{fig2} shows that this is a good approximation to the remaining fluctuation function.
Again we subtract $F^2_Y$ from $F^2_{T-X}$, change the timescale to half a year (182.625 days) by dividing both $s$ and $F$ by 18.2625 and remove points with small $s$.

So what is left if the first two components are removed from the fluctuation function? In \citep{Pla95} the authors claim to have found an 7-8 year cycle in the record of central England temperatures. Since then this mode was found in several other time series of European temperatures \citep{Jaj16}. We want to model interseasonal fluctuations with an AR(2) model
\begin{equation}
\label{AR2}
Z_{t_Z}=aZ_{t_Z-1}+bZ_{t_Z-2}+\eta_{t_Z},
\end{equation}
that - dependent on the parameters $a$, $b$ - can exhibit oscillatory behaviour with period $\tau$ (see equation (\ref{tau})) or pure relaxations. Again we treat the standard deviation $\sigma_{Z_t}$ of the process as a free parameter, which determines $\sigma_\eta$ of the noise $\eta$ according to equation (\ref{AR2V})

For Potsdam and other stations we looked at, our fit seems to show good agreement with $F^2_{T-X-Y}$, although the fluctuation function is not as smooth as for shorter times. We obtain a period of 7.5 years. The complete fluctuation function $F^2_{X+Y+Z}$ is shown in figure \ref{fig1} (right panel). It shows excellent agreement with the measured $F^2_T$.

It has been suggested that climate can be described by a continuum variability with anomalous scaling \citep{Lov15} and quasi-periodic perturbations. For the time-range under consideration, which is admittedly shorter than in most studies that concentrate on this issue, we do not see any indicators that such a model would be necessary. The scale invariance of the anomalies visible in figure \ref{fig1} (right) is explained in our model by superposition of white noise driving at each identified timescale. This is because AR(2) is a low pass filter which 'hides' the fluctuations of its driver for short times. Several AR(2) processes with different periods therefore lead to a successive increase of the background noise for longer time. This emergent scaling is an alternative interpretation to dynamical long range correlations, the origin of which is still not fully understood \citep{Fra04,Fre17,Mey17PRE}.

\section{Analysis for European station data}
\begin{figure*}
\includegraphics[width=\linewidth]{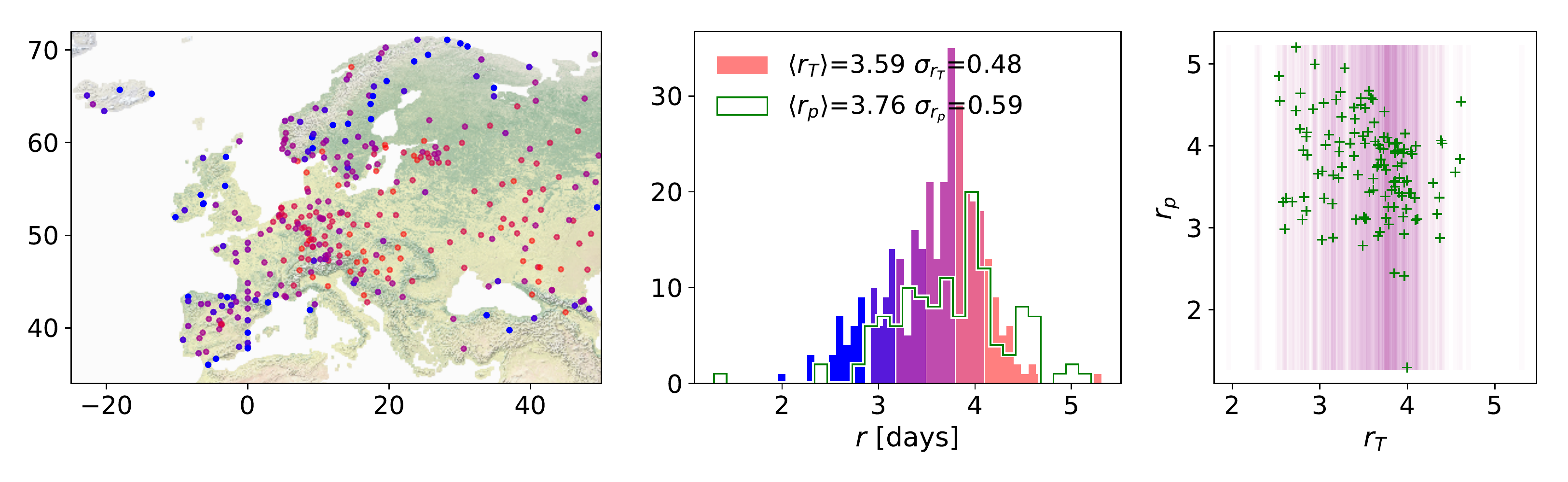}
\caption{LEFT: color plot of relaxation times $r_T$, obtained from the AR(1) parameter (\ref{AR1}) for each station. The color map can be found in the histogram in the CENTER, where it is compared with the histogram of the results of the same procedure for pressure data $r_p$ (green line). The distributions of $r_T$ and $r_p$ are similar to each other; RIGHT: $r_T$ and $r_p$ are plotted against each other (green '+') for stations where pressure data is available. The plot shows, that there is no visible correlation between both. The purple heat map in the background represents values of $r_T$ for all stations.}
\label{fig3}
\end{figure*}

\begin{figure*}
\includegraphics[width=\linewidth]{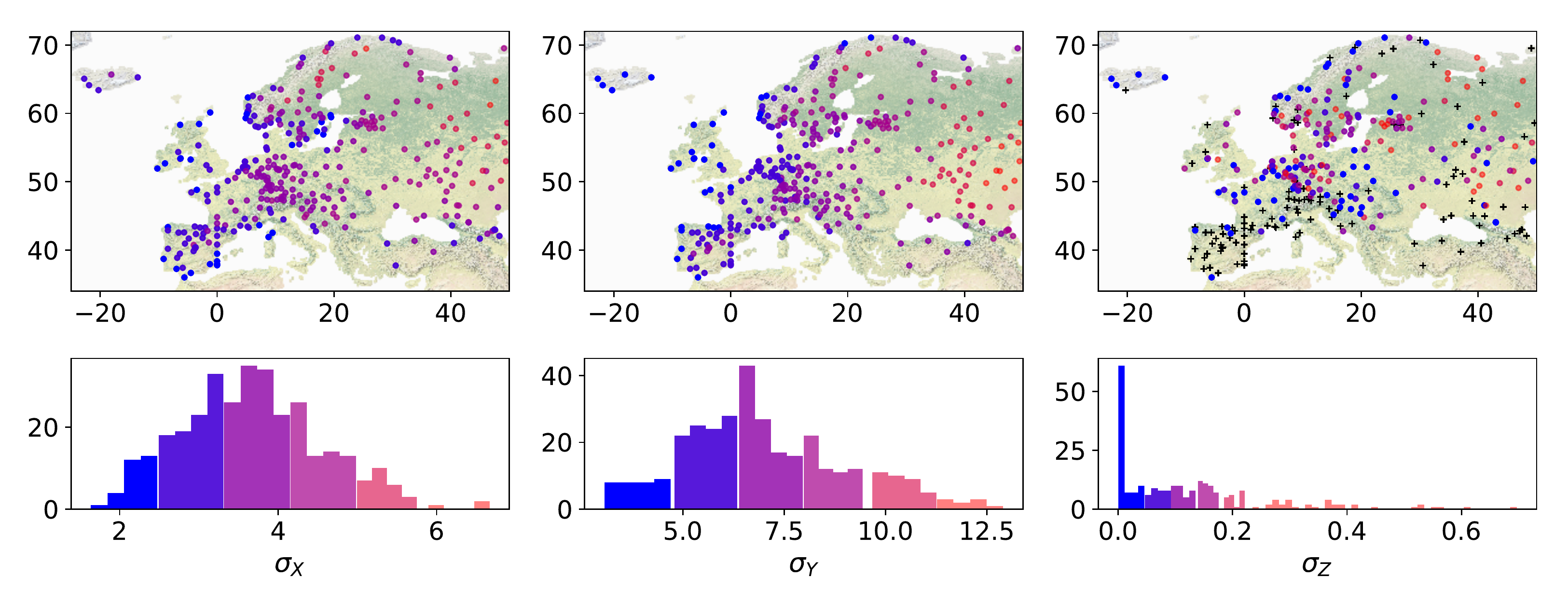}
\caption{The standard deviation in all three processes $X$, $Y$ and $Z$ represented by the parameters $\sigma_X$ (LEFT), $\sigma_Y$ (CENTER), and $\sigma_Z$ (RIGHT). For $\sigma_Z$ stations where no oscillatory behaviour in the range of 2.5 to 30 years was found are neglected and marked by black crosses. TOP: Results for all stations. BOTTOM: Histograms with identical color-coding.}
\label{fig4}
\end{figure*}

The decomposition of the fluctuation function as described in the previous section was incorporated into an algorithm and applied to temperature data  of stations around Europe. We obtain six parameters, $r$, $\sigma_X$, $\sigma_Y$, $a$, $b$, and $\sigma_Z$, for each station, which model the fluctuations of timescales up to a bit more than a decade, describing weather and macroweather. The aim is to investigate the spatial dependence of these parameters and especially for which regions the AR(2) model (\ref{AR2}) indicates an oscillatory mode on the timescale of several years. The period time $\tau$ of this oscillation can be calculated from the parameters $a$ and $b$.

The short time dynamics is described by the relaxation time $r$. We obtain it by fitting $F_X$ to $F_T$. The results are displayed in figure \ref{fig3}. $r$ shows clear regional patterns, which are much stronger than the uncertainty of our method. The correlation time is short in the west and the north of Europe close to the Atlantic. It is longer in central and eastern Europe and around the Baltic sea. For most stations $r$ is just a bit smaller than 4 days.

The dynamics of pressure data for short times turns out to be similar to that of temperature. Both describe the typical timescale of weather patterns. However, there seems to be no connection between $r_T$ and $r_p$ for one specific station. On the contrary the regional patterns of pressure strongly differ from those of temperature. Pressure values relax faster in the south compared to the north. 

The variance of the noise $\epsilon$ in the short time dynamics is very low close to the coasts and has its highest values in Russia and Scandinavia. The variances of all three processes $X$, $Y$ and $Z$ are displayed in figure \ref{fig4}. The amplitude $\sigma_Y$ of seasonality is strongly correlated with $\sigma_X$.

\begin{figure*}
\includegraphics[width=\linewidth]{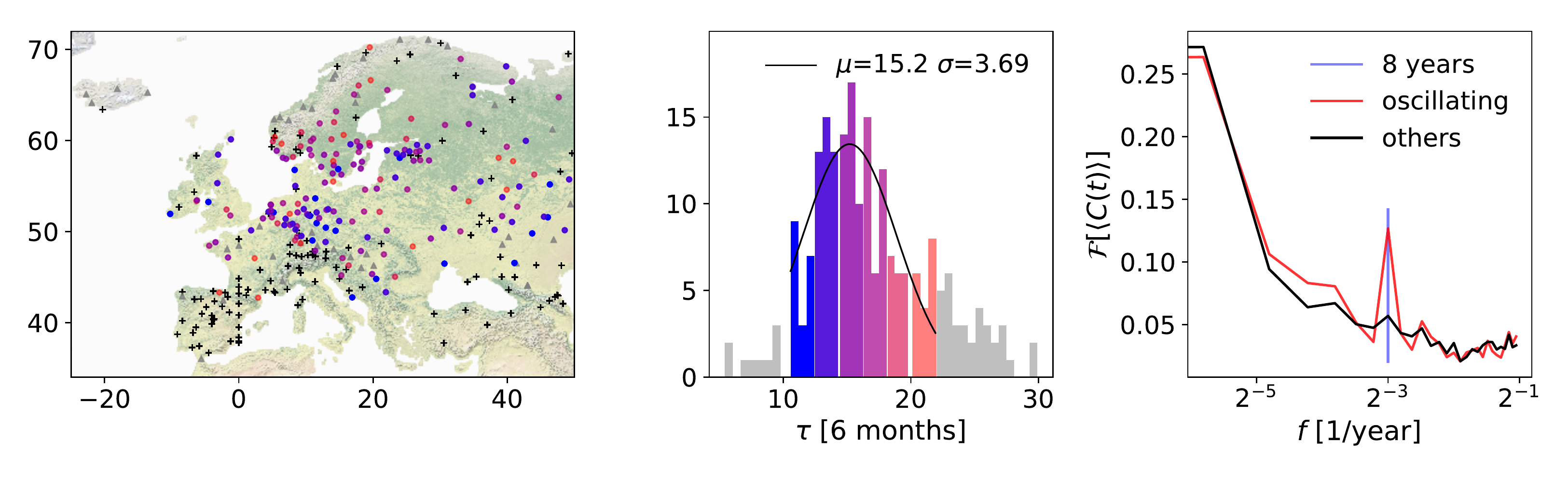}
\caption{LEFT: Results of period times $\tau$ for all stations. Fitting parameters that do not yield a period or a period far off the investigated timescale are marked as black crosses. Periods between 10 and 22 times 6 months are marked with colored dots. Other periods between 5 and 30 times 6 months are marked with grey triangles. CENTER: Histogram of colored and grey points. A Gaussian curve is fitted to the colored part via the method of least squares, yielding $7.6\pm 1.8$ years as the average period time. RIGHT: Fourier transform of correlations function averaged over all stations in $\Omega_+$ with 5 to 11 year period (red) vs Fourier transform of correlations function averaged over all other stations in $\Omega_-$ (black).}
\label{fig5}
\end{figure*}

When fitting $F_Z$ of the AR(2) model to $F_{T-X-Y}$ we only distinguish two cases. Either the model has an oscillatory mode with a period $\tau$ between 10 and 22 half years - or not. The set $\Omega_-$ of stations without this mode consists of three subsets: 
\begin{itemize}
\item if the slope is $\propto s^{0.5}$ or ever more flat the algorithm indicates an oscillation with a very short period
\item if the slope is very steep ($\propto s^{1.5}$) the algorithm indicates an oscillation with a very long period
\item for stations in between the AR(2) model has real roots indicating no oscillation.
\end{itemize}
All three cases have in common, that there is no sharp crossover in the fitted $s$-range and therefore statements about periods can only be made with high uncertainty. Examples are shown in appendix \ref{ap3}.

For 51\% of the analyzed stations we observe an oscillation, see figure \ref{fig5}. We denote the set of these stations by $\Omega_+$. The categorization into the two subgroups uncovers clear regional patterns with few exceptions which might be interpreted as false classifications. The frequency mode is significant in an area including England, Belgium, the south of Scandinavia, central Europe north of the Alps and parts of eastern Europe. The values for $\tau$ that we obtain do not show clear regional patterns, which indicates that we indeed only see the known 7-8 year cycle and the deviations are the uncertainty of our method. On average the period length is 7.6 years with large standard deviation of 1.8 years.

We test the ability of our method to distinguish between stations with and without an observable period by looking at the conditionally averaged power spectrum
\begin{equation}
P(\omega)=\mathcal{F}[\langle C(t)\rangle_\Omega],
\end{equation}
where $\mathcal{F}$ is the Fourier transform and the brackets $\langle\rangle$ denote the average over all stations in the set $\Omega_+$ or the average over all stations in the set $\Omega_-$. The validations is successful as the stations where the period was found show a clear peak at 8 years while the others do not (see figure \ref{fig5} right).

\section{Conclusions}
We introduce a method for the detection of characteristic timescales in time series based on detrended fluctuation analysis that works for data where the dominant timescales are sufficiently well separated. It is an intuitive tool for the derivation of approximate dynamics that accounts for the full power of the fluctuations in the system. In this way we bridge the gap between spectral methods that are applied for the detection of oscillations and scaling methods that detect colored noise.

While pressure data usually only shows exponential decay of correlations, temperature data shows a much slower decay, often interpreted as long range correlations. We model these time series with a simple model, a superposition of short range decay, a seasonal cycle and a potentially oscillating noisy component for longer times. This model is able to describe the observed fluctuation functions accurately. DFA is a smoothing method that is not designed to find every tiny structure in the time series. For timescales up to a decade it is not necessary to claim a colored background noise, as many authors do, who look at longer timescales. The increase of the background noise can be explained by the input noise of the interseasonal variability. By fitting the parameters we obtain the power of each component for each station. more importantly we are able to reproduce a previously found 7-8 period. Our method is with only few false classifications able to distinguish between stations where this component is significant, and stations, where it is not.

\appendix
\section{Technical details}
\label{ap3}
\begin{figure*}
\includegraphics[width=\linewidth]{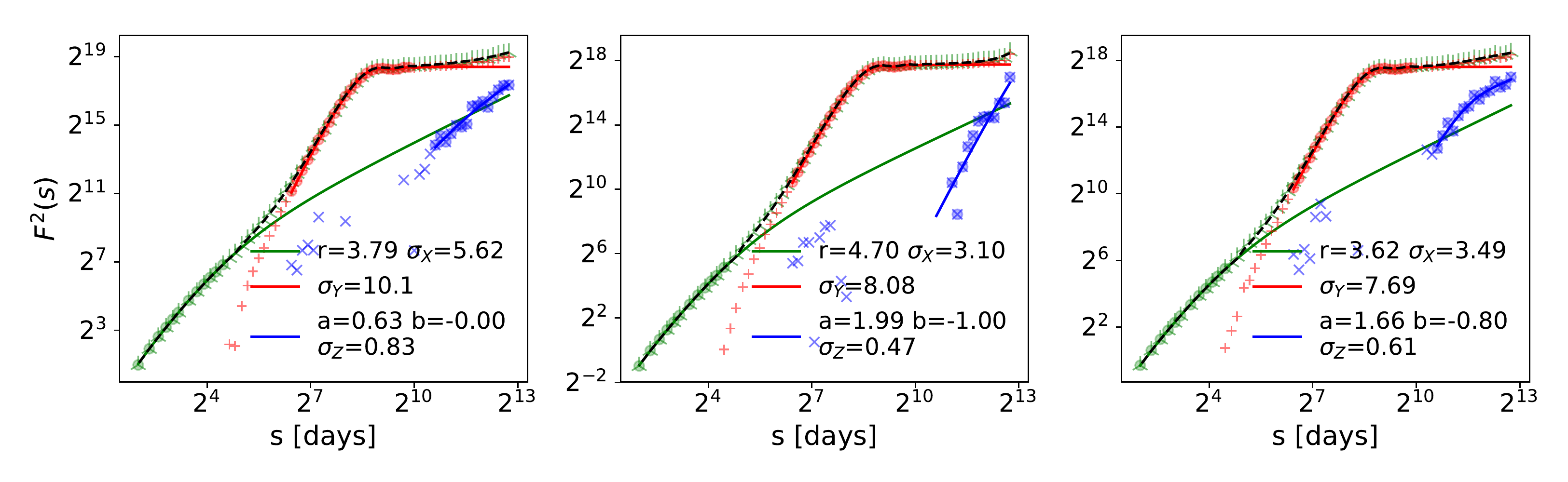}
\caption{Decomposition of $F_T$ for different stations. The color coding is identical to figure \ref{fig2} ($F^2_T$: green  '$\curlywedge$'; $F^2_{T-X}$: red '+'; $F^2_{T-X-Y}$: blue 'x'), however, here we show the full decomposition (all three steps) in one panel. The fits are drawn in the same color as the data they a fitted to. In order to increase clarity the fluctuation functions are not rescaled after each step. The dashed line in each panel represents the result $F^2_{X+Y+Z}$ of the decomposition. LEFT: Arkhangelsk, where $F^2_Z$ is best described by an AR(1) model; CENTER: Bologna, where $F^2_{T-X-Y}$ is very steep, which means that the next oscillatory mode is outside the presented $s$-range; RIGHT: Oslo Blindern, where an oscillatory mode with period length $\tau=8.2$ years was detected.}
\label{fig6}
\end{figure*}

We restrict ourselves to datasets which cover at least 60 years of (non-blended) daily temperature measurements and have at most 6.7\% of missing days. We only consider one dataset for each station in cases where there are more.
In an area around Germany with very good coverage we even require 75 years recordings. We use pressure datasets from the same stations in figure \ref{fig2} if there are at least 4 years recorded. For the analysis of the anomalies in figure \ref{fig1} we remove the 29th of February in leap years. In our main analysis, this is not necessary, since we are dealing with the real values.
In total 336 stations were considered, out of which 171 belong to $\Omega_+$, i.e. exhibit the interseasonal oscillation.

The fluctuation function is calculated for a set $S$ of $s$-values, constructed from the maximal value $s_{max}=365\cdot 75$ and the recursion $s_i=sup_{s\in \mathbb{N}} s\leq 0.9s_{i+1}$, which yields a logarithmic scale. For datasets with less than $4\cdot s_{max}$ recorded days the values of $s$ which are larger than $1/4$th of the length of the time series are ignored.

The short time behaviour of the data (\ref{AR1}) is fitted for $3\leq s\leq 25$ days. For the analysis in figure \ref{fig3}, the pressure fluctuation function is also fitted for $s\leq 25$ only, unlike in figure \ref{fig1}, where it is fitted for the whole fluctuation function as displayed.
Figure \ref{fig2} shows the full procedure for temperatures.
The seasonal cycle (\ref{SIN}) is fitted for $8< s< 38$ times 10 days.
The AR(2) model (\ref{AR2}) is fitted for $8< s$ times 6 months.
In each case fitting is done by minimizing the variance
\begin{equation}
var\left[\log\left(\frac{F_{data}(s)}{F_{model}(s)}\right)\right]
\end{equation}
The standard deviation as well as the amplitude of the seasonal cycle is not fitted but rather calculated as
\begin{equation}
\sigma=\exp\left(\sum_{s\in S} \log\left(\frac{F_{data}(s)}{F_{model}(s)}\right)\right).
\end{equation}

In figure \ref{fig6} we show the decomposition of $F^2_T$ for stations in Arkhangelsk, Bologna, and Oslo Blindern. While Oslo Blindern is another example for a station where an oscillation could be identified with a measured value of $\tau=8.2$ years, the others are examples for stations where the frequency mode is not strong enough to be significant. For Arkhangelsk our algorithm yields AR(1)-like interseasonal dynamics with a relaxation time of approximately 2.5 years. For Bologna it yields a slow oscillation due to the steep increase of the fluctuation function. The measured period length is 39 years, which is outside our investigated time range and therefore not a reliable result.

\section{Detrended fluctuation analysis}
\label{ap1}
Here we recall the
basics of DFA \citep{Pen94,Hoe15}. Given a time
series $\lbrace x_n\rbrace^N_{n=1}$ we first
calculate the integral
\begin{equation}
y_t =\sum_{j=1}^t x_j.
\end{equation}
Then we divide
the time axis into $K$ non-overlapping segments of length $s$ and calculate
the so-called DFA variance $f^2(\nu,s)$ in every segment $\nu$
which is given by the squared error sum of $y_t$ and a fitting polynomial $p_t$ of order $q$. This order gives the name of the method DFAq
\begin{equation}
f^2(\nu ,s)=\frac{1}{s}\sum_{t=1+(\nu-1)s}^{\nu s} (y_t-p_t)^2.
\end{equation}
We repeat the procedure for an other $K$ non-overlapping segments of length $s$, but this time starting from the end of the series.
Finally, the square of the DFA fluctuation function is the
average of all the DFA variances over all segments
\begin{equation}
{F}^2(s)=\frac{1}{K}\sum_{\nu=1}^{2K} f^2(\nu,s).
\label{segmentssum}
\end{equation}
Traditionally people look at the asymptotic behaviour
${F}^2(s)\sim s^{2\alpha}$, which contains information about the correlation structure. For $\alpha = 1/2$ the process is short range correlated and for $\alpha > 1/2$ the process is long range correlated.

\section{Fluctuation functions of AR(1) and AR(2)}
\label{ap2}
Theoretical fluctuation functions can be calculated from equation (\ref{F(C)})
\begin{equation}
{F}^2(s)=\sigma^2\left(L_q(0,s)+2\sum_{t=1}^{s-1}C(t)L_q(t,s)\right).
\end{equation}
as proposed in \citep{Hoe15a} and performed in \citep{Hoe15} and \citep{Mey19NJP}. 
The kernel $L_1$ in DFA-1 is
\begin{equation}\begin{array}{ll}
L_1(t,s)&= \frac{1}{30(s^4-s^2)}[3t^5-5(4s^2-1)t^3+30(s^3-s)t^2\\
&-(15s^4-35s^2+8)t+2(s^5-5s^3+4s)].
\end{array}\end{equation}
The correlation function of AR(1) is known to be
\begin{equation}
\label{AR1C}
C(t)=c^t,
\end{equation}
with AR(1) parameter $c$.
The fluctuation function for AR(1) can therefore be calculated as
\begin{equation}
\label{Far1}
{F}_c^2(s) = \sigma^2\frac{c^s J_c(s) + K_c(s)}{15(c-1)^6(s^2-s^4)},
\end{equation}
with $J_c(s)$, $K_c(s)$ polynomials in $s$
\begin{equation}\begin{array}{ll}
J_c(s) &=60[s^2(c^2-c)^2-3s(c^3-c)+2(c^4+c^3+c^2)],\\
K_c(s) &= s^5(c-1)^5(c+1)+15s^4c(c-1)^4\\
&-5s^3(c-1)^3(1-7c-7c^2+c^3)\\
&-15s^2c(c-1)^2(1-10c+c^2)\\
&+2s(c-1)^3(2-17c-17c^2+2c^3)\\
&-120c^2(1+c+c^2).
\end{array}\end{equation}
The variance of the process is given by
\begin{equation}
\sigma^2={\sigma_\eta^2}/{(1-c^2)},
\label{AR1V}
\end{equation}
where $\sigma_\eta^2$ is the variance of the noise.

For AR(2) the correlation function reads
\begin{equation}
\label{CAR2}
C(t)=h_1g_1^t+h_2g_2^t,
\end{equation}
where $g_1$ and $g_2$ are real or complex roots obtained from rewriting the definition (\ref{AR2}) as
\begin{equation}
(1-g_1B)(1-g_2B)x_t=\eta_t,
\end{equation}
with the backshift operator $BX_t=X_{t-1}$.
The constants $h_1$ and $h_2$ are calculated from $C(1)={a}/{(1-b)}$ and $C(0)=1$.
The model might exhibit oscillatory dynamics with a period
\begin{equation}
\tau=\frac{2\pi}{\arctan(\mathrm{Im}(g_1)/\mathrm{Re}(g_1))}
\label{tau}
\end{equation}
if $g_1$ and $g_2$ are complex.
Its variance is
\begin{equation}
\label{AR2V}
\sigma^2=\frac{(1-b)\sigma_\eta^2}{(1+b)(1-a-b)(1+a-b)}
\end{equation}
The fluctuation function for AR(2) can be expressed in terms of the fluctuation function of AR(1), because of the correlation function (\ref{CAR2}) being the superposition of two AR(1) fluctuation functions. However, different variances of the two AR(1) processes have to be taken into account
\begin{equation}
\label{Far2}
{F}_{a,b}^2(s)=\sigma^2 \frac{[c_1{F}_{g_1}^2(1-g_1^2)+c_2{F}_{g_2}^2(1-g_2^2)](1-b)}{(1+b)(1-a-b)(1+a-b)}.
\end{equation}
Here ${F}_{g_i}$ are the fluctuation functions of the AR(1) processes with parameters $g_1$ and $g_2$.

\begin{acknowledgments}
We acknowledge the data providers in the ECA\&D project.
Klein Tank, A.M.G. and Coauthors, 2002. Daily dataset of 20th-century surface air
temperature and precipitation series for the European Climate Assessment. Int. J. of Climatol.,
22, 1441-1453.
Data and metadata available at http://www.ecad.eu.
Map background made with Natural Earth. Free vector and raster map data @ naturalearthdata.com.
\end{acknowledgments}

\bibliographystyle{natbib}

\end{document}